\documentclass[prl,twocolumn,graphicx,amssymb,floatfix]{revtex4}

\usepackage{graphicx}
\begin{document}

\title{Reply to the Comment ``Past of a quantum particle and weak measurement''. }

\author{L. Vaidman}
\affiliation{ Raymond and Beverly Sackler School of Physics and Astronomy\\
 Tel-Aviv University, Tel-Aviv 69978, Israel}

\begin{abstract}
Misinterpretation  in the preceding Comment of my recent analysis of the past of a photon is corrected. There is nothing in this analysis which is ``contrary to the usual quantum expectations'' but, nevertheless, it does provide ``further understanding and interpretation of the system considered''. In particular, it teaches us that the naive common sense argument frequently used in the literature including the Comment has to be abandoned.    \end{abstract}
\maketitle

First I want to spell out the results of the Comment \cite{Li}  which support the results of my paper \cite{past} contrary to the claim in the Comment.
``A knowledge [measurement] of the weak value partially destroys the quantum interference in the nested Mach-Zehnder interferometer.'' The authors of the Comment correctly construct and analyze a model of such weak measurement in Section III.  They are right that  disturbance of the interference due to weak measurement explains the results presented in my paper in the framework of the standard approach. I provided the same explanation with less details but mentioning the key issues in the last paragraph of Sec V: ``In the standard formalism of quantum mechanics it can be explained as a counterintuitive interference effect. ...''

Let us turn now to the results with which I disagree. The authors of the Comment write: ``the weak value is not a directly observable quantity in any real experiment''. Tenths of experiments directly measured weak values of observables. Both real and imaginary parts of the weak value were observed: the shift of the pointer variable of a standard measuring device coupled weakly to the observable yields the real part and the shift of the conjugate to the pointer variable yields the imaginary part \cite{AV90}. In most cases, and in particular in the case in question, these shifts are much smaller than the uncertainty of the pointer, so we need a measurement on an ensemble, but it does not make these measurements ``indirect''.

I disagree that ``the weak value only tells us the level that the original system is perturbed''. Weak value is defined by pre- and post-selection. It can be revealed (on a large enough ensemble) with as little perturbation as required. I do not find any meaning in the remark: ``... the weak value itself is not ``weak'',..''.

The authors develop an alternative meaning of the concept of weak value through the meaning of the denominator and the numerator of the weak value expression. I, as one of the inventors of  the concept of the weak value, consider it as an effective value of an observable of a pre- and post-selected quantum system in the limit of a weak coupling to that observable. The weak value expression or its ingredients appear in various physical formulae, but it does not change the meaning of the concept. Also, I could not follow the derivation in the Comment. The authors write: ``The quantity
$|\langle \psi_f | |001\rangle \langle 001| | \psi_i \rangle |^2$ represent[s] the probability of $D_1$ clicking under the condition that the photon is found at
the position C.'' It seems to me, however, that this conditional probability does not depend on the initial state.

The Comment's main point is that the results of my paper have an alternative simple explanation without the two-state vector formalism. I  see a little merit in their ``rough answer''  why the weak values of projections on B and C do not vanish while the weak value of the projection on E vanishes. But anyway, without the backward evolving state this argument fails to explain why  the weak value of the projection on F vanishes too.

I disagree with the way the authors of the Comment apply ``the logic of weak values'', calculating in Eq. (13) the ``joint weak value''. What they calculate is the weak value of the product of the projection on C and the projection on E. The weak value of the projection on E is zero, but the product rule fails for weak values of pre- and post-selected systems \cite{product}, so the weak value of the product needs not be zero. What is relevant are separate weak values of the projection on C and the projection on E. These values can be measured together, i.e. simultaneously but with separate measuring devices, with a vanishing disturbance in the limit of a large ensemble.

My paper shows that a ``common sense'' argument applied in the first section of the Comment, ``... it is reasonable to assume that the single photon must have followed the outer path A and the probability of its existence inside the smaller Mach-Zehnder interferometer (along paths F, B, C, and E) must be zero", has to be abandoned. Therefore, the counterfactuality of the recent protocols \cite{CF1,CF2} which relies on this argument, fails. But this is an issue discussed elsewhere \cite{CFV1,CFV2}.

Let me conclude with a clarification of the quotation in the Comment from  my other paper \cite{CFV1}:``The photon did not enter the interferometer, the photon never left the interferometer, but it was there''. One might run into a paradox understanding it in a naive sense according to which if ``it was there'' it could not be simultaneously in any other place. Photon is a quantum particle and in this particular case it was also in the other arm of the large interferometer outside the interferometer of the quotation. The meaning of ``the photon was there'' is that it left the weak trace there. And it left no weak trace on the paths towards and from the interferometer.

What apparently led the authors to write their Comment is that, strictly speaking, the last sentence of the preceding paragraph is true only in the limit of an ideal experiment, but then, there is also no trace inside the interferometer. The justification of my claim: ``the photon did not enter the interferometer and the photon never left the interferometer'' is that the ratio between the amplitude of the trace in these outer paths of the interferometer and the amplitude of the trace in the inner paths of the interferometer goes to zero in the limit of weak measurement. The ratio of the trace inside the interferometer and the trace in the other arm of the large interferometer remains to be 1 in that limit.  Since everywhere there are nonzero tails of quantum wave functions, we should not ask for an exact zero amplitude to say that ``the photon was not there''.

This work has been supported in part by  grant number 32/08 of the Binational Science Foundation and the Israel Science Foundation  Grant No. 1125/10.

\end{document}